\documentclass[twocolumn,preprintnumbers,amsmath,amssymb,nofootinbib
,superscriptaddress]{revtex4}
\usepackage{hyperref}
\usepackage{graphicx}
\usepackage{color}
\usepackage{xcolor}

\newcommand{\be}{\begin{equation}}  
\newcommand{\ee}{\end{equation}}  
\newcommand{\bear}{\begin{eqnarray}}  
\newcommand{\eear}{\end{eqnarray}}  
\newcommand{\ba}{\begin{array}}  
\newcommand{\ea}{\end{array}}

\usepackage{diagbox}

\definecolor{rossoCP3}{cmyk}{0,.88,.77,.40}
\definecolor{blueRef}{rgb}{0.2,0.2,0.6}
\definecolor{blue}{rgb}{0,0.396,0.741}
\hypersetup{
	colorlinks, 
	bookmarksopen, 
	bookmarksnumbered,
	citecolor=blueRef, 		
	linkcolor=rossoCP3,	
	urlcolor=rossoCP3,			
}

\newskip\humongous \humongous=0pt plus 1000pt minus 1000pt

\newif\ifdtup

\def\oldreffmt#1{\rlap{[#1]} \hbox to 2\parindent{}}

\def\figfmt#1{\rlap{Figure {#1}} \hbox to 1in{}}  
  
\def\etal{\hbox{\it et al.}}  
  
\def\Tr{\mathop{\rm Tr}}

\def\slash#1{#1\!\!\!/\!\,\,}  
\def\beq{\begin{equation}}  
\def\eeq{\end{equation}}  
\def\bea{\begin{eqnarray}}  
\def\eea{\end{eqnarray}}  
\def\half{\frac{1}{2}}  
  
\def\bq{\begin{quote}}  
\def\eq{\end{quote}}

\def\half{\frac{1}{2}}       

\def \etal {{\it et al.}\ }  
\relax

\newdimen\tdim  
\tdim=\unitlength  
\def\bar{\overline}

\begin{document}
\preprint{FERMILAB-PUB-21-757-T}

\title{Naturally Self-Tuned  Low Mass Composite Scalars \\ 
}

\author{Christopher T. Hill}
\email{hill@fnal.gov}
\affiliation{Theoretical Physics Department, \\
Fermi National Accelerator Laboratory, \\
P. O. Box 500, Batavia, IL 60510, USA}

\begin{abstract}

Scalar bosons composed of a pair of
chiral fermions in a non-confining potential 
have an effective Yukawa coupling, $g$, to free external chiral fermions. 
At large distance a Feynman loop
of external fermions generates a scale invariant potential, $V_{loop}\propto -g^2/{r^{2}}$,
which acts on valence fermions for separation $\rho=2r$.  This generally
forces the $s$-wave ground state to deform to a static,  zero mass, configuration,  
and for slowly running, perturbative $g$, 
a large external ``shroud'' wave-function forms. 
This is related to old
results of Landau and Lifshitz in quantum mechanics.
The 
massless composite scalar boson ground state is then an extended object.
Infra-red effects can generate a small mass for the system. This
points to a perturbative BEH-boson composed of top and anti-top quarks and
a novel dynamical mechanism for
spontaneous electroweak symmetry breaking. 
\end{abstract}

\maketitle
 
\date{\today}

\email{hill@fnal.gov}


\section{Introduction}

For approximately fifty years particle physics has dealt with a conundrum: 
The electroweak hierarchy problem, the apparent unnaturaness 
of low mass scalar particles,  or,
why is the Brout-Englert-Higgs (BEH-boson) mass, or weak scale, small compared to e.g., the Planck scale?  
This has driven much of the thematic research for half a century, 
from supersymmetry \cite{SUSY},  technicolor \cite{FS} and extended technicolor \cite{ETC}, 
top condensation \cite{Yama,BHL,Topcolor,NSD}, ``composite models'' 
(where the BEH-boson is a pseudo--Nambu-Goldstone
mode \cite{Georgi}), etc. 
The discovery
of the BEH-boson in 2012 at the LHC, and the apparent lack of
any nearby new physics to act as a custodian, has exacerbated the conundrum.  The BEH-boson
appears to be, for all practical purposes, an approximately massless (e.g., on the Planck scale) 
scalar field. This is seemingly anathema to fifty years of post-modern theoretical physics.

In the present paper we look more closely at the internal dynamics of
bound states consisting of chiral fermions in non-confining potentials.  We will show
that approximate scale invariance, 
in conjunction with chiral symmetry,
manifests itself in an unusual way in bound states
and leads to unexpected consequences for 
composite solutions.

Here we will see that a scalar boson can form as a compact massive object,  
a ``core'' wave-function,
consisting of a pair of chiral fermions, bound by some
short-distance interaction potential. 
This can happen 
at an arbitrarily high mass scale, $M$, potentially as 
high as $M\sim M_{Planck}$, and one usually assumes this
state cannot then have a naturally small mass, $m<<M$.  
However, if the potential is not confining, then chiral and scale symmetries conspire 
through a Feynman loop, external to the core, to create a large-distance,
attractive, scale invariant, $-cg^2/r^2$ potential between the constituents,
where $r$ is the radius of the two-body system and $c$ is a loop factor.
This is the ${\cal{O}}(\hbar)$ vacuum reaction to the presence of the core, and
it is an effect usually phrased in momentum space
that is central to the Nambu--Jona-Lasino model \cite{Nambu}.

The constituent fermions are virtually emitted and reabsorbed into the core, experiencing the vacuum
effects, as
in Fig.(1). The  induced vacuum potential leads to 
an enveloping ``shroud'' wave-function around the core. The shroud
is necessarily a massless solution owing to the scale invariance of the vacuum potential.
However, for a consistent solution its  null
time dependence 
must match, via boundary conditons, onto the core
wave-function. 

This happens by a deformation of the short-distance core, locking it 
to a static, zero mass configuration.
Indeed, if one allows arbitrary boundary conditions there are
generally massless solutions for any core potential, but
these don't become eigenfunctions because they are matched to exterior solutions
in a normal vacuum, typically radiation. This then yields the large mass, approximate eigenvalue.
With the vacuum loop potential the core wave-function can deform 
and match onto the exterior massless shroud solution. The full solution becomes
an eigenfunction with a zero mass eigenvalue.  We exhibit this explicitly
in a simple model, but it is a general phenomenon. Due to scale symmetry,
the shroud wave-function is an extended object, and the low energy 
physics becomes insensitive to the core. 

At first this seemed surprising to us, but after arriving at
this conclusion and the relevant wave-function of the shroud, we found there is a 
prior (somewhat obscure) discussion of related effects 
in the immortal ``{\em Nonrelativistic Quantum Mechanics}'' 
textbook of Landau and Lifshitz \cite{LL} (LL).
They explored the Schroedinger equation in  $-\beta/r^2$ potentials
and found the extended wave-functions that apply to our present situation, though the context and
some details differ.
Moreover, they argued that the existence 
of a zero energy ground state is guaranteed in any core potential 
(modulo any negative energy modes) if the  $-\beta/r^2$  is present at large distances.

Quoting from Landau and Lifshitz,  page 116, 
of the edition, \cite{LL} (we insert our comments in {\em italics} and for us ``energy''
becomes $M^2$):
\begin{quote}
``Next, let us investigate the properties of the solutions 
of Schroedinger's equation in a field which 
diminishes as $U=-\beta/r^2$, 
and has any form at short distances. 
({\em For weak coupling}) it is
easy to see that in this case only a finite number of negative energy
levels can exist. For with energy $E=0$ 
Schroedinger's equation at large distances has the form (35.1)  with the
general solution (35.4) {\em(our eqns.(\ref{ours},\ref{ours2})}. 
...

Finally, let the field be $U=-\beta/r^2$ in all space.  Then for ({\em weak coupling})
there are no negative energy levels. For the wave-function of the state $E=0$
is of the form 
(35.7) in all space; ... i.e., it corresponds to
the lowest energy level."
\end{quote}
This is essentially the statement that the shroud solution controls
the entire solution, i.e. ``the tail wags the dog,'' and in a not-so ``weak coupling'' 
limit, $g^2 < {8\pi^2}/{N_c},$ (see eq.(\ref{49})) this can generally 
 be the ground state of the system with $M^2=0$. 

The LL solutions provide a potential new mechanism for achieving light composite 
scalar bosons, based upon internal dynamics and symmetries.
We believe this may provide a candidate solution to 
the electroweak hierachy problem and the structure of the BEH-boson,
though our present discussion is confined to a single complex scalar field
with global chiral covariance.
The LHC may be seeing the ``shroud'' 
of the ground state solution, the extended
structure of the BEH-boson.

The spatial extent of the shroud is cut-off when the 
chiral symmetry of the constituents is broken, which may be triggered by other forces.
In this picture, if the BEH-boson is composed of top and anti-top quarks, 
it would have an extent
of order $r \sim 1/m_{top}$. The renormalization
group (RG) running of the top Yukawa BEH-coupling may act perturbatively as the trigger for electroweak
symmetry breaking. 
Essentially, we view this in reverse: the top quark gets a mass, which cuts off
the shroud solution. Owing to the minus sign of the vacuum loop
potential, this leaves a tachyonic  mass term (``Mexican hat'' potential)
for the composite BEH-boson. This in turn causes its vacuum expectation 
value (VEV) to form, which 
in turn generates the top quark mass. The
self consistency determines the critical value of $g=g_{top}$
at which this occurs, and we indeed find $g\approx 1$.

 There are
requisite stability constraints, e.g., no negative $M^2$ solution at the short distance core
scale is allowed, and the BEH-Yukawa coupling must not run too quickly,
e.g., near a quasi-fixed point of the RG to
obtain the shroud solution over a large range of scales.
The top quark BEH-Yukawa coupling obliges the latter and the exclusion
of negative $M^2$ follows from
weak dynamics, such as barrier potentials, new non-confining
gauge interactions, and possibly gravitation.  We
emphasize that {\em this is not a strong dynamical theory, 
and works perturbatively with $g\sim O(1)$}.

If the BEH-boson is  an extended object it
would behave coherently as a pointlike particle
at LHC processes probed thus far, but perhaps its compositeness 
can be seen
in higher energy or sensitive flavor processes, or perhaps in deep
$s$-channel production in a muon collider \cite{muon}.
These issues will not be discussed in the present paper.

We believe, however, that this may be pointing to an intimate
relationship between three quantities: the BEH-boson mass of $125$ GeV,
the top quark mass of $175$ GeV, and the VEV of the BEH-boson
$246$ GeV (or $175$ GeV when divided by $\sqrt{2}$).
We sketch a trigger mechanism for the
spontaneous breaking of the $SU(2)\times U(1)$ symmetry,
coming from the QCD contribution to the RG running
of the top-quark BEH-Yukawa coupling.

After a discussion of formalism and a ``warm-up'' example in Section II,
we derive the relevant Landau-Lifshitz solutions and construct 
low mass scalar bound states in Section III.  In Section IV we discuss
infrared mass and normalization, and sketch a theory of the origin
of the electroweak scale, the top quark mass and BEH-boson mass.
We conclude in Section V, and present detailed loop calculations,
particularly of the vacuum loop potential, 
in Appendix I.

\section{Composite Systems}

\subsection{Hints from the NJL Model}

Many years ago Ken Wilson demonstrated how to solve the Nambu-Jona--Lasinio
 model, (NJL) \cite{Nambu}, in a conceptually powerful
 way by the renormalization group \cite{Wilson}.
The NJL model is the simplest field
theory of a composite scalar boson,
consisting of a pair of chiral fermions. The chiral fermions induce loop effects
that lead to the interesting dynamical phenomena at low energies \cite{Wilson,BHL}.

Consider a pair of chiral fermions, 
\bea
\left( \psi _{R}^{a},\psi
_{L}^{b}\right) 
\eea
with $N_{c\text{ }}$color indices $(a,b)$,
and a global chiral symmetry $U(1)_{L}\times U(1)_{R}.$
The NJL model with its non-confining, local, chirally invariant
interaction takes the form:
\bea
\label{NJL11}
L=\frac{g^{2}}{M^{2}}(\overline{\psi }_{L}^{a}\psi _{aR})(\overline{\psi }%
_{R}^{b}\psi _{L,b})+L_{kinetic}
\eea
(we'll henceforth suppress summed color indices). 

We factorize this by introducing an auxilliary
field $\Phi $:
\bea 
\label{factor}
L_{M }=g[\overline{\psi }_{R}{\psi}_{L} ]\Phi +h.c -M^{2}\Phi ^{\dagger }\Phi 
\eea
Integrating out $\Phi$ in eq.(\ref{factor}) we recover eq.(\ref{NJL11}).

Wilson viewed this as the effective action at a scale $M.$  He then computed
fermion loop
corrections that arise because the chiral fermions are unconfined
and wander into the vacuum. 
This yields the theory at a lower mass scale $\mu$.
\bea
\label{NJLM}
L_{M }&\rightarrow & L_{\mu }= g[\bar{\psi} _{R}\psi_L ]\Phi + h.c -V_M\Phi ^{\dagger }\Phi+...
\nonumber \\
\!\!\!\!\!\makebox{where,}&& 
V_M = \!\left(\!M^{2}\!-\!\frac{N_{c}g^{2}}{8\pi ^{2}}\left( M^{2}-\mu ^{2}\right)\right)
\eea
Here ... includes an induced kinetic term and quartic interaction 
which we computed in a large $N_c$
fermion loop approximation \cite{BHL,CTH} :
\bea
\label{five}
&& 
\!\!\! Z_HDH^{\dagger }DH-\frac{\lambda }{2}\left( H^{\dagger
}H\right) ^{2};\;\;\;
Z_H=c_{1}\!+\!\frac{N_{c}g^{2}}{16\pi ^{2}}\ln \left(\! 
\frac{M^2}{\mu^2 }\!\right) \nonumber \\
&&
\lambda =c_{2}+\frac{2N_{c}g^{4}}{16\pi ^{2}}%
\ln \left(\! \frac{M^2}{\mu^2 }\!\right).
\eea
The log terms give the leading large $N_{c}$ fermion loop corrections to the
kinetic and quartic terms, and yield  a running of the couplings,
e.g., $ g\sim 1/\sqrt{Z_H}$,
which can be
matched onto the full RG  equations in the IR \cite{BHL,Yama,CTH}. 
Indeed, the arguments of the logs inform us that the RG is operant on all
scales, $\mu$ to $M$.
We recover these results in the pointlike limit of our composite field
discussion in Appendix I, and they are largely retained
when one looks at RG running in $r$. 
For further pedagogical discussions see
\cite{CTH}).

Note, in particular, the behavior of the composite scalar boson mass in
$V_M$ of eq.(\ref{NJLM}).
The $ -{N_{c}g^{2}M^{2}}/{8\pi ^{2}}$  term arises from the negative
quadratic divergence in the loop involving the
pair $\left( {\psi }_{R},\psi _{L}\right) $ of Fig.(1), with
pointlike vertices and a loop cut-off scale at $M^{2}$. 
This is the physical response of
the vacuum to the classical interaction $g[\overline{\psi }_{R}\psi
_{L}]\Phi $ in the presence of the bound state $\Phi $.
The Dirac sea generates a
feedback to reduce $M^{2},$ and the loop integral is then capturing
this physical effect, much like a Casimir effect.

The NJL model allows us in principle to fine-tune the coupling $g^{2}$ to a critical
value,
$g_{c}^{2}={8\pi ^{2}}/{N_{c}}$, at which point the mass of the bound state becomes zero. 
In the earliest models of  a composite BEH-boson, known as ``top
condensation,''
\cite{Yama,BHL}, we tuned the theory to have a massless, or slightly
supercritical, bound state, by ``human intervention.''   Note
there is a hint of
something special about the critical value, since this corresponds to a
cancellation
of the large $M^{2}$ terms in the theory, and an approximate scale symmetry emerges, broken
only by log terms and the infrared cut-off,  $\mu^2$.  

Fine--tuning done by human intervention cannot be
viewed as a complete or satisfactory theory.  To generate
a hierarchy
where $M/\mu \sim M_{Planck}/v_{weak}\sim 10^{17}$ requires tuning $g^{2}$ to $%
g_{c}^{2}$ with a precision of \ $1:10^{-34}.$
This graphically illustrates the electroweak hierarchy problem. 
Nevertheless, the NJL model informs us that
composite scalar bosons, consisting of a pair of chiral fermions
with a non-confining potential, can indeed exist and will have an induced 
or fundamental Yukawa coupling $g$.

\subsection{Self-Tuning }

In a realistic model with more detailed binding dynamics, however, the 
possibility of an emergent scale symmetry
suggests that the NJL fine-tuning cancellation may 
actually be a ``self-tuning'' effect.
The  internal wave-function of the
bound state might adjust itself to find a new
ground state which possesses the maximal scale invariance. After all,
the NJL model is an effective field theory and only captures physics on IR
scales $ \mu <<M$,
but is blind to the detailed internal dynamics,
requiring we probe deeper.

The main observation in the present paper is that,
viewed in configuration space, external chiral fermions induce
an extended, scale invariant, attractive loop potential for the bound state 
wave-function of the form $-cg^2/{r}^2$. 
This  particular potential has the nontrivial zero mass solutions
of Landau and Lifshitz (LL) \cite{LL}.
This then leads to the ``self-tuning'' where 
the short distance part of the solution becomes
 locked to the LL exterior solution.  
 
 We will also see below that there
 is an intimate connection between the NJL model and the LL solutions
 as they share an identical  ``critical coupling,'' even though the former
 case is controlled by a quantum loop while the latter is a classical result.
 To us this bolsters the relevance of the LL solution for
 non-confined bound states of chiral fermions.

We interpret the  ``custodial symmetry'' of the massless system 
to be the approximate scale invariance of the $-1/{r}^2$ potential
modulo soft RG running of couplings.
Various IR effects can subsequently generate a natural small mass
for the composite system.
This is a self-consistent  phenomenon
since the valence constituents of the bound state experience a potential
due to virtual effects of the same particles in the vacuum via a Feynman loop.
It is similar in this sense to a Coleman-Weinberg potential \cite{CW} in which quantum
fluctuations of a field $\phi$ induce a potential for the VEV of $\phi$.

In this picture,
the composite scalar boson becomes an extended object. 
This is evidently the price one pays for naturalness.
In top condensation models we had assumed a pointlike BEH-boson
bound state \cite{BHL}, but we were forced to fine-tuning. Presently,
we allow the theory, via the vacuum loop potential, to relax the bound state and
we obtain the shroud as an extended object. 
Now we do not require fine tuning, and we can
have perturbative coupling.

We can examine the log-terms of eq.(\ref{five}) in configuration
space and see that
the usual  renormalization group behavior
of the BEH-Yukawa coupling is evidently retained
as logarithmic functions of the scale $r$, and the LL solution
will be maintained if $g$ is approximately constant,
i.e. an approximate RG fixed point.
The induced potential creates a quasi-conformal window with
the wave-function extending from the UV scale
of the short-distance binding, to the large IR scale 
of the mass generation.  Hence a hierarchy is dynamically generated.

Any mass for the shroud  requires explicit IR modification of the $-1/{r}^2$ potential. 
The infrared scalar boson
mass can be treated explicitly and it is technically natural when inserted by hand
where the potential becomes $\rightarrow m^2-1/{r}^2$.
However, we expect this will be generated dynamically, e.g.,
through the IR behavior of the Yukawa coupling, or a Coleman-Weinberg
mechanism \cite{CW} generates mass through the running of the quartic coupling \cite{HillCW}.
This mechanism is general and offers various model
building possibilities.  

As one possibility, we
outline a simple, self-consistent origin of the top quark and BEH-boson masses.
The main result here is that the LL solution provides a natural,  massless scalar field,
due to the inner conformal window.

\subsection{Formalism for Composite Fields}

Consider a hypothetical new fundamental interaction associated
with a high energy scale, $M$:
\bea
\label{TC}
L'=g_0^2 [\bar{\psi}_{L}(x) \gamma_\mu T^A \psi_L(x)] \; D(x-y)\;
[\bar{\psi}_{R}(y) \gamma^\mu T^A \psi_{R}(y)]
\nonumber \\ 
\!\!\!\!\!
\eea
where $T^A$ are generators of an $SU(N)$ interaction
and the $\psi$ fields are in the fundamental representation, e.g.
color triplets for $SU(N_c=3)$ of color. 
This is a broken gauge theory with massive gluons, analogous
to ``topcolor,'' \cite{Topcolor}, however {\em we will not
require that this be a strongly interacting theory}, i.e.,
$g_0$ need not be large. 

A Fierz rearrangement of the interaction
leads to:
\beq
 L'\rightarrow  -g_0^2[\bar{\psi}_L(x)\psi_{R}(y)] D(x-y)[\bar{\psi}_{R}(y)\psi_L(x)]
+ O(1/N_c)
\eeq
where combinations of fields in the $[...]$ are color contracted.
We can now factorize this into an effective interaction
with a bilocal auxilliary field:
\bea
\label{eight}
L'&\rightarrow & g_0[\bar{\psi}_L(x)\psi_{R}(y)]\Phi(x,y)+ h.c. 
\nonumber \\
&& - \;\Phi^\dagger(x,y)D^{-1}(x-y)\Phi(x,y).
\eea
Note that, apart from normalization, this is a bilocal generalization of
eq.(\ref{factor}).

\begin{figure}
	\centering
	\includegraphics[width=.15\textwidth]{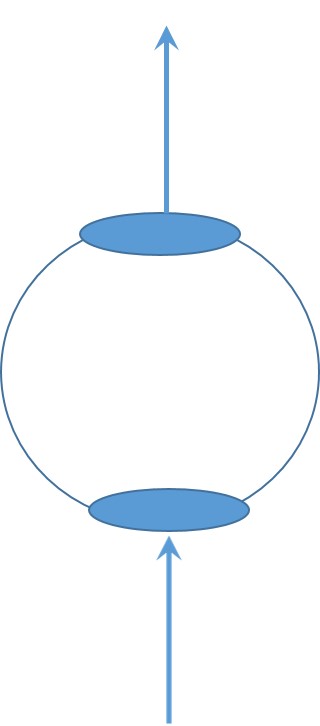}
	\caption{ Fermion loop with wave-function $\phi(r)$ vertices 
	which generates the vacuum loop potential term in the action,
	$-\eta\phi^2(r)/r^2$.} 
	\label{fig:loop}
\end{figure}

We go to a space-like hyper-surface
and impose that the constiuent fermions share a single common time coordinate
in this frame.  
The bilocal composite field takes the form,
$\Phi (\vec{X},\vec{r},t)$ where $\vec{X}$ is the center-of
mass coordinate, $\vec{\rho}=2\vec{r}$ the interparticle separation.\footnote{
The single time constraint  can be made manifestly Lorentz invariant, e.g.
the invariant condition $P_\mu r^\mu=0$ with the 4-momenum $P_\mu$
and $r^\mu = (x^\mu-y^\mu)/2$.
A bilocal invariant action is then ``gauge fixed'' on a time slice
as $S= \int d^4\rho d^4X \delta(\sqrt{P_\mu\rho^\mu})=P_0^{-1}
\int d^3\rho d^4X$.
Fully covariantized expressions rapidly become awkward,
as they would be for any conventional composite system, such as proton, atom or molecule.
We will work in the rest frame knowing the results can always be boosted with care.}

In the rest frame we have  the single time variable $t=X^0$. 
 $\Phi (\vec{X},\vec{r},t)$ may be
viewed as a ``bosonization'' of the s-wave component of the fermion operator
product on time slice $t$,
\bea 
&&
[\overline{\psi}_{R}\left(\vec{X}-\vec{r},t\right)\psi_{L}\left( \vec{X}+\vec{r},t\right)] 
\rightarrow \Phi (\vec{X},\vec{r},t)+..
\nonumber \\
&& \;\;\;\; X^0=t,\qquad \vec{X} = \frac{\vec{x}+\vec{y}}{2},
\qquad \vec{r}=\frac{ \vec{x}-\vec{y} }{2}.
\eea
Presently we will ignore gauge interactions and focus on a singe complex bound state field.
We  factorize $\Phi$ as:
\bea
\Phi(x,y)=\chi(X^\mu)\phi(\vec{r})\qquad (t,\vec{X})=X^\mu
\eea
where the time dependence is carried by the pointlike
factor, $\chi(X^\mu)$, and
$\phi(\vec{r})$ is then a static ``internal wave-function.''
Typically $\chi(X)\sim \exp(iP_\mu X^\mu)$ describes the motion
of the center of mass of the system, such as a plane wave, and $\phi$ describes the bound state
structure and dynamics. The action is a
function of  $\Phi$, and 
such terms as, e.g.,  $|\chi^4\phi^2|$, $|\phi^2|$,  etc., are disallowed.

 The single
time dependence can be carried by either field, though typically we choose $\chi(t) $
hence it is a quantum field with canonical dimension of mass.
 $\phi(\vec{r})$, on the otherhand,  is typically static, and we presently treat it classically. As 
 a static field $\phi$ has no canonical momentum and 
satisfies a static differential equation.
Since we are working with a classical $\phi$ we will assume it is dimensionless.
It forms a static ``configuration,''
something like an instanton.
The full composite field $\Phi=\chi\phi$ is canonical and in the pointlike
limit, $\phi\rightarrow \delta^3(\vec{r})$, $\Phi$ becomes a local quantum field, essentially
pure $\chi(X^\mu)$.

We will exclusively consider
a ground state composed of a pair of fermions in an $s$-wave, so $\phi(\vec{r})$ is  
spherically symmetric under rotations of the radius $\vec{r}=\vec{\rho}/2$ with $\vec{X}$
held fixed.
Hence the bilocal interaction term of eq.(\ref{eight})
yields the action, including an assumed central potential, $V_{0}(\vec{r})$, and
kinetic terms, one for the center-of-mass
and the other for the radius:  
\bea
 \label{action}
S\!=&&\!\!\!\!\!\!\!\!\int \!\!\frac{ d^3 r}{\hat{V}} d^4\!X\! \left( 
 \!Z_\chi\! | \phi^2||\partial_\chi{\chi}|^2\!\!- \!\!Z_\phi|\chi|^2| \partial_r{\phi }|^2
\!\!\! -\!V_{0}(\vec{r})|\chi\phi |^{2}\right)
\nonumber \\
& &\!\!\!\!\!\!\!\!\!\!\!\!\!\!\!\!\!\!
-g_0\!\!\int \!\!  \frac{ d^3r}{\hat{V}} d^4X [\bar{\psi}_L(\!\vec{X}+{\vec{r}})
\psi_{R}(\!\vec{X}-{\vec{r}})]\chi(\!\vec{X})\phi(\vec{r})\!+\!h.c.
\eea
Here $
\vec{x} =\vec{X}+\vec{r}$ and $\vec{y} =\vec{X}-\vec{r}$
and we have,
\bear
\partial_x^2+\partial_y^2 = \half \partial_X^2 +  \half \partial_r^2
\eear
Hence we have a ``bare'' relationship $Z_\chi =Z_\phi= 1/2$. 

We have introduced a ``normalization volume,'' $\hat{V}$,
to maintain canonical dimensionality of the overall space-time action integral.
No physical quantites depend upon $\hat{V}$.
Note that with the normalization condition,
\bea
\label{norm0}
\int  \frac{d^3 r}{\hat{V}}\phi^2(r) =1.
\eea
 and with a rescaling $\chi\rightarrow \chi/\sqrt{Z_\chi}$ we can make the $\chi$ kinetic term canonical.
A renormalized action is then:
\bea
 \label{actionr}
S\!=&&\!\!\!\!\!\!\!\!\int \!\!\frac{ d^3 r}{\hat{V}} d^4\!X\! \left( 
  | \phi^2||\partial_\chi{\chi}|^2\!\!- \!\!z|\chi|^2| \partial_r{\phi }|^2
\!\!\! -\!V_{r}(\vec{r})|\chi^2\phi^{2}|\right)
\nonumber \\
& &\!\!\!\!\!\!\!\!\!\!\!\!\!\!\!\!\!\!\!\!\!\!
-g\!\!\int \!\!  \frac{ d^3r}{\hat{V}} d^4X [\bar{\psi}_L(\vec{X}\!+{\vec{r}})
\psi_{R}(\vec{X}\!-{\vec{r}})]\chi(\vec{X})\phi(\vec{r})+h.c.
\eea
where,
\bear
g=Z_\chi^{-1/2}g_0\qquad\;\;\; V_r\rightarrow Z_\chi^{-1}V_0
\eea
 In general, we have the parameter 
$z=Z_\phi/Z_\chi$ which is potentially subject to radiative corrections.
To simplify the present discussion we will adopt the value $z=1$.

A global $U_L(1)\times U_R(1)$ chiral symmetry is now the $U(1)$ transformation 
$\Phi \rightarrow e^{i\theta} \Phi$. We presently ignore a potentially
thorny  issue of local gauge covariance,
which requires internal Wilson lines.

Working in the 
 rest-frame, $\chi\propto e^{iMt}$, 
the mass $M$ of the bound state is then determined by the eigenvalue of the static
equation for the ground state in $\phi$ (see the next subsection, II.D):
\bea
\label{eigen}
\left( \frac{d^{2}}{dr^{2}}+\frac{2}{r}\frac{d}{dr}-V_{r}(r)\right) \phi
(r)=-M^{2}\phi (r)
\eea
Here $V_{r}(r)$ is the core potential that binds the fermions into
quasi-stable, approximate eigenstates, but is not confining.
Substituting this into the  action, eq.(\ref{action})
where we integrate over $r$ and apply eq.(\ref{norm0})
we obtain an effective point-like 
action for $\Phi(X,r)\rightarrow \Phi(X)$, 
 \bea
 \label{action3}
 S&=&  
\!\! \int d^4X\;\big( | \partial{\Phi}|^2-M^2|\Phi|^2\big)
 \nonumber \\
 && \!\!\!\!\!\!\!\!\!\!\!\!\!\!
-g\!\int \!\! d^4X [\bar{\psi}_L(X)\psi_{R}(X)]\Phi(X)+h.c.+...
\eea
The ellipsis, $...$, includes corrections to $g$ from an expansion in $r$, involving higher derivatives,
$\sim r^{n}\partial^{n}\psi(X)$, which may provide probes of compositeness.

 Note we will require that the relevant solutions to the equation of motion for physical
bound states at short distances, $r\sim M^{-1}$
must have a real mass eigenvalue $M$, hence $M^{2}\geq 0$.  A negative $%
M^{2}$  represents a vacuum
instability at short distances.
For the barrier potential below we can enforce this by positivity of $%
V_{r}(r)$.

\subsection{Warm-up: A Simple Composite Model With Barrier Potential }

We now consider a simple barrier potential model,
which
leads to a straightforward textbook quantum mechanics
problem.  This illustrates the bosonized formalism and the derivation
of the BEH-Yukawa coupling in a general potential model, where it
may not be present {\em ab initio}, as in eq.(\ref{action}).
The present warm-up model ignores the induced vacuum loop potential and
the Landau-Lifshitz solutions.

We will presently assume the  renormalized action
and we'll neglect the $g$, Yukawa term.
Varying $\chi$,
from eq.(\ref{actionr}) with $z=1$, it follows that: 
\bea
&& \!\!\! \!\!\! \!\!\! \partial^2\chi \!\int \!\!\frac{ d^3r}{\hat{V}} | \phi^2|
=\chi\!\!\int \!\!\frac{ d^3r}{\hat{V}} |\left(-| \nabla_r{\phi }|^2
\!\!\! -\!V_{r}(r)|\phi |^{2}\right)
\eea
We assume an eigenvalue, $M^2$, and varying with respect to $\phi$,
we then have the separate equations of motion:
\bea
\label{eqmo}
&&\partial^2 \chi(X) =-M^2\chi(X),
\\
&&\left( \frac{d^{2}}{dr^{2}}+\frac{2}{r}\frac{d}{dr}-V_{r}(r)\right) \phi
(r)=-M^{2}\phi (r).
\eea
Note the $\chi$ equation is a free Klein-Gordon form, while the
$\phi$ equation is static.

To simplify we can work in the rest frame,  and
impose the normalization conditions:
\bea
\label{norm1}
\int \!\!\frac{ d^3r}{\hat{V}} | \phi^2|=1 \qquad \chi=\frac{1}{\sqrt{2MV}}\exp(iMt)
\eea
where $\chi$ has a conventional plane wave ``box normalization'' where $\hat{V}$ 
is an imaginary volume associated with the internal $3$-space, and
$V$ is the volume of an imaginary exterior $3$-space box (these volume factors cancel
in physical quantites).
Note that the $\chi$ terms become canonical in eq.(\ref{action})
with the above $\phi$ normalization.

Consider a ``thin wall'' barrier potential:
\bea
\label{barrier1}
\makebox{Region I:} &\qquad&  V_r(r<R)=0  \nonumber \\
\makebox{Region II:} &\qquad&   V_r(R<{r}<R+a)=W^{2}\nonumber \\
\makebox{Region III:} &\qquad&   V_r(r>R+a)= 0
\eea
In contrast to nonrelativistic quantum mechanics where 
the barrier has dimensions of energy, here the barrier, $W^2$, 
has dimensions of (mass)$^2$.

The general solution for eq.(\ref{eqmo}) is:
\bea
\makebox{Region I:} &\;\;&  \phi (r) = {\cal{N}}\frac{\sin (k{r})}{r};\qquad  k=M  \nonumber \\
\makebox{Region II:} &\;\;&  \phi (r)={\cal{N}}'\frac{e^{-\kappa{r}}}{r};\qquad 
\kappa =\sqrt{W^{2}-k^{2}} \nonumber \\
\makebox{Region III:} &\;\;\;& \phi (r)= a \frac{e^{iM{r}}}{r}+b 
\frac{e^{-iM{r}}}{r}
\eea
Region III is pair radiation, since for $\Phi=\chi\phi$, we  have:
\bea
\label{chisoln}
\!\!\!\!
\makebox{Region III:} &\;\;&
\Phi= a\chi_0 \frac{e^{iM(t+r)}}{r}+b \chi_0
\frac{e^{iM(t-r)}}{r}
\eea
a sum of incoming (left-moving) and outgoing (right-moving) spherical waves.
If we set $a=0$ we have the outgoing $s$-wave of a fermion pair from
the decay of the bound state.

The matching of Region I to Region II requires:
\bea 
\tan (kR)&=&-\frac{k}{\kappa },\;\;\;\;\; {\cal{N}}^{\prime }={\cal{N}}e^{\kappa
R}\sin (kR)
\eea
Note that, as usual, the boundary matching conditions determine $k$  
and the eigenvalue,  $M=k$.  For large $\kappa$
we have $kR\rightarrow \pi$ and $\sin(kR)\rightarrow -k/\kappa$,
$\cos(kR)\rightarrow 1$.

The matching of Region II to Region III requires:
\bea
a&=&\frac{1}{2}{\cal{N}}\left( 1+i\frac{\kappa }{k}\right) \sin (kR)e^{-ik\left(
R+a\right) -\kappa a}\;\;\nonumber \\
b&=&  \frac{1}{2}{\cal{N}}\left( 1-i\frac{\kappa }{k}\right) \sin
(kR)e^{ik\left( R+a\right) -\kappa a}
\eea
The normalization integral of eq.(\ref{norm1}) is dominated by the cavity Region I
and yields approximately, with $kR=MR=\pi$:
\bea
\label{norm2}
&&
1=\int \!\!\frac{ d^3r}{\hat{V}} | \phi^2|
=4\pi \int_0^R \!\!\!{\cal{N}}^2\sin^2(k{r}) \frac{dr}{\hat{V}}
=\frac{2\pi^2{\cal{N}}^2}{M\hat{V}}
\nonumber \\
&&
{\cal{N}}^2=\frac{M\hat{V}}{2\pi^2}.
\eea
(note, the Region II contribution to the mass, in the large $\kappa$ 
and thin wall $a<<R$ limit, is
negligible
$\sim M^{2}a+O(aM/\kappa ).$)

The solution represents a steady state, a balance of an incoming
and outgoing radiative part.  It cannot be matched to a pure outgoing 
wave unless the core solution explicitly decays in time, which then requires integrals over
Green's functions.
However, if we are interested in an initial state, consisting of one pair
of fermions localized in the
Regions I+II, then we can switch off the incoming
radiation, $a\rightarrow 0$, and the state will decay, where the 
decay amplitude is $b.$  
The decay width is obtained semi-classically by the rate of energy
loss (power) into the outgoing spherical wave, divided by the mass.

Here's a cursory semi-classical {\it estimate} of the decay width.
In the rest-frame with no explicit dependence upon $\vec{X}$
we see that $\Phi$ can be viewed as a localized field in $\vec{r}$ and $X^0=t$, hence
$\Phi(r,t)=\phi(r)e^{iMt}/\sqrt{2MV}$.
The outgoing power is given by the stress tensor, $T_{0r}$,
from the right-mover solution. Note that $\kappa \sin(kR)/k\rightarrow -1$
in the large $\kappa$ limit and 
\bea
|\phi(r)|^2\sim |b|^2/r^2\sim {\cal{N}}^2e^{-2\kappa a}/4r^2,
\eea
Hence the power, using eq.(\ref{norm2}):
\bea
P &=&  
\!\!\int\!\! d^3 X\frac{4\pi r^{2}}{\hat{V}} \left( \partial _{0}\Phi ^{\ast }
\partial_{r}\Phi  \right)\big|_{r\rightarrow \infty}
\nonumber \\
&\approx&  \pi M^2 V \frac{{\cal{N}}^2}{(2MV\hat{V})} e^{-2\kappa a}=
\frac{1}{4\pi} M^2e^{-2\kappa a}
\eea
and the decay width is obtained as the ratio,
\bea
\Gamma &=&\frac{P}{M}= \frac{M}{4\pi}e^{-2\kappa a}
\eea
We can compare the decay width from a complex pointlike
field consisting on a single color $N_c=1$, of mass $M$ with Yukawa coupling $g$
to the fermions:
\bea
\Gamma =\frac{%
g^{2}}{16\pi }M
\eea
Matching this to the composite model calculation gives
\bea
g=2e^{-\kappa a}
\eea
We therefore have a heavy bound state with mass $M=k\approx \pi /R$
and an induced Yukawa coupling $g\propto e^{-\kappa a}$ which  is
perturbative
in the large $\kappa a\sim Wa$ limit. 

We've done this for a single color. In this simple model if
we extend to $N_c$ colors, then $\Phi$ will receive a color
normalization factor of $1/\sqrt{N_c}$ and the mass will then
become $M \times N_c/N_c$ unchanged. The decay width we have computed
semiclassically is also unchanged as color sums cancel against
this normalization factor.
When we compare to the field theory decay width with $N_c$
colors, we see that our model predicts
$g_f=g/\sqrt{N_c}$, and yields a color suppressed decay.
The above calculation assumed $g_0=0$ (zero bare Yukawa coupling)
and obtained the induced effective coupling $g\sim 1/\sqrt{N_c}$.
However, the coupling $g$ need not be induced,
and can come directly from a fundamental $g$ and is then ${\cal{O}}(1)$
rather than $1/\sqrt{N_c}$. 

The main takeaway is that the eigenvalue $M^2$ is  generated by the matching of Regions I, II and the
radiative Region III.  It is the matching that determines the eigenvalue and
dictates the relevant solutions to the differential equations in each region.

\newpage

\section{Fermion Induced Vacuum Loop Potential}

\subsection{Discussion }


The full action for $\Phi$, including only $V_0(r)$, is incomplete.
Since there is a Yukawa coupling to the exterior fermions, either fundamental
or induced,
we must include the feedback effect arising from the last term
in eq.(\ref{action}) of integrating out
fermion fields, as in eq.(\ref{NJLM}). 
The chiral fermions
roam through the surrounding space and affect the vacuum.
The Feynman loop. of Fig.(1) 
takes the form of an attractive,
approximately scale invariant ``vacuum loop potential''
which we denote as $V_{loop}(r)$.
This can be seen by direct calculation in Appendix I:
\bea
\label{loop1}
 V_{Loop}(r)=-\frac{\eta}{{r}^{2}}, \qquad \eta = \frac{%
N_{c}g^{2}}{32\pi^2}
\eea 
Note that ${r}$ is radial and not a Compton wavelength, 
hence its associated momentum is $1/{r}$. 

We can intuit the form of eq.(\ref{loop1}) by comparing to the 
 momentum space form of the loop, the ${\cal{O}}(\hbar)$
 term in $V_M$  of eq.(\ref{NJLM}),
 \bea
V_M = M^2-  \hbar\frac{N_{c}g^{2}}{8\pi ^{2}}\left( M^{2}-\mu ^{2}\right)
 \eea
 At large distances,
$ {r}\sim L>>R$,
the bound state will acquire mass, which provides an IR cut-off
on the potential in the Lagrangian,
\beq
V_{Loop}(r) \sim - \left(\frac{\eta}{{r}^2} - \frac{\eta}{L^2} \right)
\eeq
This matches the sign of the  $\mu^2\sim 1/4L^{2}$ term in $V_M$.
 Likewise, the short-distance behavior $\sim -\frac{\eta}{{r}^2}$
 with $r^2\sim 1/4M^2$
matches the $-g_0^2N_cM^2/8\pi^2$ term in  $V_M$.

The key feature for us is that $V_{Loop}(r)$ contains no mass scales if
$g^{2}$ is constant, i.e., if $g^{2}$ is an approximate fixed point of the RG evolution
into the IR.  
This means that there is a region outside of $V(r)$, such as Region III 
in our previous example, in which the 
potential is the scale invariant $V_{Loop}$, and
in this region the solution is scale invariant, with $M=0$.

These are the solutions studied by Landau and Lifschitz in \cite{LL}.

\subsection{Scale Invariant Landau-Lifshitz Solutions}

We assume that $V_{Loop}$ grows more negative until the scale of the core radius
of the bound state $r =R$ is reached, then becoming a
constant negative vacuum energy in the core, $-{\eta}/{R^2 }$ for ${r}<R$.
We 
will presently assume an additional constant core potential, $ -U^2$, 
so that  $V_r(r<R) = -U^2-{\eta}/{R^2 }$.

Remarkably we can omit $U^2$ altogether and we
still have binding from the vacuum loop potential alone.  That is, if we simply pinch a pair of
chiral fermions together they will generate a self-binding potential and a nontrivial
self-consistent solution, hence we expect that even a comparatively
weak force, such a gravity, can trigger the formation of these states.

We therefore  have:
\bea
\makebox{Region I:}&& \;\; V_r(r<R)=-U^2 -\frac{\eta}{R^{2}}\nonumber \\
\makebox{Region II:} && \; \;V_{loop}(r>R)=-\frac{\eta}{{r}^{2}}
\eea
Assuming $\eta$ is approximately constant we find for the form of the Region II solutions, following LL.
The scale invariant spatial differential equation eq.({\ref{eqmo}) becomes,
\bea
\label{ours}
\left( \frac{d^{2}}{dr^{2}}+\frac{2}{r}\frac{d}{dr}+\frac{\eta }{{r}^{2}}\right) \phi
(r)=-M^{2}\phi (r)
\eea
Following Landau and Lifshitz, 
this is solved with the anzatz $\phi (r)=r^{p}$, \cite{LL},
to obtain:
\bea
\label{ours2}
p^{2}+p+\eta =0,  \qquad M^2 =0
\eea 
hence, we find:
\bea
\label{ours3}
p_{1}&=&-\frac{1}{2}+\frac{1}{2}\sqrt{\left( 1-4\eta \right) }\nonumber \\
p_{2}&=& -\frac{1}{2}-\frac{1}{2}\sqrt{\left( 1-4\eta \right) }
\eea
The solution has  $M^2=0$, hence these are static massless solutions.
 
We see that there are two distinct
cases:  $\eta < 1/4$ and  $\eta > 1/4$.   
Our main interest will be in the weak coupling case, $\eta < 1/4$.
 Remarkably, $4\eta>1$ and $4\eta<1$, which
emerge from the solutions to the classical differential equation,
anticipate the critical
behavior of the NJL model, which arises from the loop integral. That is,
 $4\eta=g_0^2N_c/8\pi^2 =1$  corresponds to the critical value,
 and $\eta < 1/4$ ($\eta > 1/4$) is an unbroken phase (broken phase)
 of the NJL model, as seen in eq.(\ref{NJLM}).
The classical LL solutions thus anticipate the
critical coupling and distinct phases of the NJL model!

In Region I, for any core potential $V_r(r)$, we can generally find a static solution as well.
Presently we take,
\bea
\phi (r)&=& {\cal{N}}\frac{\sin (k{r})}{r}
\eea
and find that with the choice,
\beq
\label{k}
k^{2}=U^2+\frac{\eta }{R^{2}}
\ee
we have a zero mass $M^2=0$.  Note that $U^2$ can have any sign and magnitude,
and if $k^2<0$  the $\sin (k{r})\rightarrow \sinh (|k|r)$.

It is of key importance to note that 
 $k$ is now determined by eq.(\ref{k}) alone,
and not by the matching boundary condition of I to II to radiation III, as in our previous example. 
Since this is possible for any potential (we can slice any potential into multiple segments with
different values of $U^2$, and match at each segment boundary)
and the scale invariant solution will always exist \cite{LL}.  There may be
negative $M^2$ solutions which are model dependent but must be disallowed.
These are disllowed here for negative $U^2$ where the potential in I+II
resembles a ``castle with moat.''  

The full solution is then: 
\bea
\!\!\!\!\!\!\!\!\makebox{Region I:}\;(r<R)\;\;\; \phi (r)&=& {\cal{N}}\frac{\sin (k{r})}{r}\qquad\nonumber  \\
& &   k^{2}=U^2+\frac{\eta }{R^{2}}%
\qquad   \ \  
\nonumber \\
\!\!\!\!\!\!\!\!\makebox{Region II:}\;(r>R)\;\;\; \phi (r)&=&\frac{A}{R}\left( \frac{r}{R}\right) ^{p_{1}}+\frac{B}{R}\left( 
\frac{r}{R}\right) ^{p_{2}}
\nonumber \\
\eea
where  matching of Region {\bf I} to Region {\bf II} requires, 
\bea  {\cal{N}}\sin (\sqrt{\beta })&=&A+B
\nonumber \\
{\cal{N}}\sqrt{\beta }\cos (\sqrt{\beta })& =& A\left( \ 1+p_{1}\right) +B\left(
1+p_{2}\right) 
\eea
where:
\bea
kR=\sqrt{\beta }
\eea
hence, 
\bea
A&=&\frac{\cal{N}}{\left(  p_{2}-p_{1}\right) ^{-1}}\left( \left( \ 1+p_{2}\right) \sin (%
\sqrt{\beta })-\sqrt{\beta }\cos (\sqrt{\beta })\right) 
\nonumber \\
B&=&-\frac{\cal{N}}{\left(  p_{2}-p_{1}\right) ^{-1}}\left( \left( \ 1+p_{1}\right) \sin (%
\sqrt{\beta })-\sqrt{\beta }\cos (\sqrt{\beta })\right) 
\nonumber \\
&&
\eea
In the case of strong coupling we have  a
complex expression, since $\eta >\frac{1}{4}$,
\bea
p_{1}&=&-\frac{1}{2}+\frac{1}{2}\sqrt{\left( 1-4\eta \right) }=-\frac{1}{2}+%
\frac{1}{2}i\xi 
\nonumber \\
p_{2}&=&-\frac{1}{2}-\frac{1}{2}\sqrt{\left( 1-4\eta \right) }=-\frac{1}{2}-%
\frac{1}{2}i\xi 
\eea
where $\xi =\left| \sqrt{\left( 1-4\eta \right) }\right| $.
Hence,
\bea
&&\!\!\!\!\!\!\!\!\!\!A=\!\!i\xi^{-1}\!{\cal{N}}\!\left( \frac{1}{2}\left( 1-i\xi \right) \sin (%
\sqrt{\beta })-\sqrt{\beta }\cos (\sqrt{\beta })\right) 
\nonumber \\
&&\!\!\!\!\!\!\!\!\!\!B=\!\!-i\xi^{-1}\!\!{\cal{N}}\!\left( \frac{1}{2}\left( 1+i\xi \right) \sin (%
\sqrt{\beta })-\sqrt{\beta }\cos (\sqrt{\beta })\right) 
\eea
and the general solution is:
\bea
\phi (r)&=&\frac{|A|e^{i\sigma}}{R}\left[\left( \frac{r}{R}\right)^{(-1+i\xi)/2}
\!\!+\left( 
\frac{r}{R}\right) ^{(-1-i\xi)/2}\right]
\nonumber \\
e^{2i\sigma}&=&-\frac{\left( 1-i\xi \right) \sin (%
\sqrt{\beta })-2\sqrt{\beta }\cos (\sqrt{\beta }) }{ \left( 1+i\xi \right) \sin (%
\sqrt{\beta })-2\sqrt{\beta }\cos (\sqrt{\beta }))}
\eea

The case where $U=0$ we have  $\beta = \eta$.
The solution  in the weak coupling limit,
$\eta <\frac{1}{4},$  is a simpler real expression:
\bea
\label{49}
p_{1}&=&-\frac{1}{2}+\frac{1}{2}\sqrt{\left( 1-4\eta \right) }\approx
\allowbreak -\eta +O\left( \eta ^{2}\right) 
\nonumber \\
p_{2}&=&-\frac{1}{2}-\frac{1}{2}\sqrt{\left( 1-4\eta \right) }\approx
\allowbreak \allowbreak -1+\eta +O\left( \eta ^{2}\right) 
\eea
and,
\bea
A&=& -\left( 1-2\eta \right) ^{-1}{\cal{N}}\left( \left( \eta \right) \sin (%
\sqrt{\eta })-\sqrt{\eta }\cos (\sqrt{\eta })\right)
\nonumber \\
&\approx &
 {\cal{N}}\sqrt{\eta }+\frac{1}{2}{\cal{N}}\eta ^{\frac{3}{2}}+O\left( \eta ^{\frac{5}{2}%
}\right) 
\nonumber \\
B&=& \left( 1-2\eta \right) ^{-1}{\cal{N}}\left( \left( \ 1-\eta \right)
\sin (\sqrt{\eta })-\sqrt{\eta }\cos (\sqrt{\eta })\right) 
\nonumber \\
&\approx&
-\frac{2}{3}{\cal{N}}\eta ^{\frac{3}{2}}+O\left( \eta ^{2}\right) 
\eea
hence,
\bea  
\phi (r)&=&\frac{A}{R}\left( \frac{r}{R}\right) ^{p_{1}}+\frac{B}{R}\left( 
\frac{r}{R}\right) ^{p_{2}}
\eea
or for small $\eta$.
\bea
\phi (r)&=&\frac{{\cal{N}}\sqrt{\eta}}{R}\left( \frac{r}{R}\right)^{-\eta}
\!\!-\frac{2{\cal{N}}{\eta^{3/2}}}{3R}\left( 
\frac{r}{R}\right)^{-1+\eta}
\eea
Note in the real case the 
$B$ term falls off faster in the IR
while the magnitude of both $A$ and $B$ is the same
in the complex case at large distance.

\subsection{ Barrier Potential with $V_{Loop}$ and LL Solutions}

As a simple example of a full solution with core and shroud, 
we return to the barrier potential and include the presence of a nonzero $g$.
Therefore we must
match onto the
LL solution in Region III, rather than onto the radiative solution.
Here we assume the potential $V_r$ as defined in eqs.(\ref{barrier1}) for the Regions I and II.
This is actually ``unrealistic'' in the sense that we are ignoring
the additional $V_{loop}$ effect for Region I, $r<R+a$. However, this
shows the generality of the matching effect with the tunneling barrier in Region II.  

\begin{figure}
	\centering
	\includegraphics[width=0.5\textwidth]{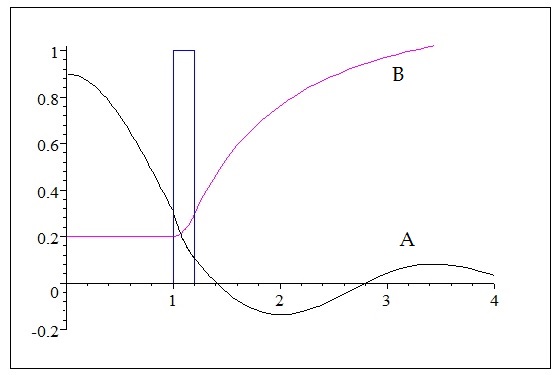}
	\caption{
	Barrier potential solutions: (A) solution with only the barrier, $V_r(r)$, and
	${\cal{N}}=0.4$, $U^2=25$, $k=2.6$, hence $\kappa=\sqrt{W^2-k^2}=4.27$;
	(B) exact solution with barrier $V_r(r <R+a)$ and  vacuum loop potential, 
	$V_{Loop}(r>R+a) = -\eta/r^2$,
	with $\phi_0=0.5$, $\eta = 0.2$, $R=1$,  $a=0.2$, $\kappa=5$.
	Neither solution is normalized. Note how solution (A) displays a lump in the core
	that matches onto radiation external to the barrier, while (B) has a flattened core with zero mass
	to match onto the LL solution ``shroud'' in the exterior.
	}
	\label{fig:loop}
\end{figure}

Consider a ```castle and moat'' barrier potential:
\bea
\makebox{Region I:} &\qquad&  V_r(r<R)=0  \nonumber \\
\makebox{Region II:} &\qquad&   V_r(R<{r}<R+a)=W^2 \nonumber \\
\makebox{Region III:} &\qquad &   V_{loop}(r>R+a)= -\frac{\eta}{r^2}
\eea
Here we have chosen $U^2 +{\eta}/{R^{2}}=0$ (negative $U^2$) for simplicity 
and comparison to the
previous ``warm-up'' example of a
barrier potential with $V_r=0$ in its Region I;
Any value of $U^2$ will produce an LL solution in Region III.

The solution in the three regions is now, with $\kappa=|W|$:
\bea
\makebox{Region I:} &\;\;&  \phi (r) = \phi_0 \nonumber \\
\makebox{Region II:} &\;\;&  \phi (r)= \frac{\phi_0}{2r \kappa}\times\nonumber \\
&& 
\left((1+\kappa R)e^{\kappa(r-R)}-(1-\kappa R)e^{-\kappa(r-R)}\right)
\nonumber \\
\makebox{Region III:} &\;\;\;& \phi (r) =\frac{A}{R}\left(\frac{r}{R}\right)^{p_1}
+\frac{B}{R}\left(\frac{r}{R}\right)^{p_2}
\eea
where $A$ and $B$ are rather messy expressions which we quote in the
limit $a/R<<1$ and $\eta <<1$:
\bea
\frac{A}{R}&=&\phi_0\frac{(\kappa R\eta-1)\cosh(\kappa a)+(\eta-\kappa^2R^2)\sinh(\kappa a)}{\kappa R(2\eta-1)}
\nonumber \\
\frac{B}{R}&=&\phi_0\frac{\kappa R\eta\cosh(\kappa a)+(\eta+\kappa^2R^2-1)\sinh(\kappa a)}{\kappa R(2\eta-1)}
\nonumber \\
\eea

The solution is shown as (B), with comparison to the
``warm-up'' example (A), in Fig.(2).
First we note that Region I is the solution (B) with free boundary conditions is
$\phi=\phi_0=$ constant.
In the solution of Fig.(2) (A), which is the barrier potential
of II.C, the boundary matching conditions determined $k$  
and the eigenvalue,  $M=k$.  On the other hand for (B)
the Region I solution is a trivial constant, $k=0$.
This reflects that the scale invariance in this case tends to ``flatten'' the core
wave-function.  
The matching of Region I to Region II then requires
that there are both exponentially increasing and decreasing components
in the barrier.

In Region III in Fig.(2) we see that the previous solution (A) with $\eta=0$
matches onto radiation, 
while now (B) with $\eta$ nonzero
matches onto the LL solution. This grows with $r$
to a maximum value then
attenuates like $(r/R)^{p_1}\sim r^{-\eta}$
as $r\rightarrow \infty$.

\newpage

\section{ IR Mass and Normalization}

We emphasize that the LL solutions will always force the overall bound state solutions to be
massless.  The internal wave-function $\phi(r)$ presently satisfies
a linear differential equation and can be freely renormalized (though
we will contemplate a quartic interaction below).
However, one sees that the massless LL solutions
are not compact and are, without an IR cut-off, non-normalizeable.
We require an IR cut-off to the solution, which we will
define to be $L$. This can come from an IR mass term and
leads to $L\sim 1/m$,

We also see that these solutions are very insensitive to the core
structure, and nearly vanish as $r\rightarrow R$.
We will presently focus on the dominant LL solution in the IR
in the small $\eta$ limit which is the $A$ component.
Given the core insensitivity
it is inconvenient to maintain explicit dependence upon $R$. 
Hence we will renormalize the solution and use an IR cutoff $L$ as the unique scale,
\bea
\phi(r)\rightarrow
\phi_r\left( \frac{r}{L}\right)^{-\eta}
\eea

It is now convenient to redefine the  normalization:
\bea
&&\int \frac{ d^3r}{\hat{V}} \; \phi^2
= \int_0^L \frac{4\pi r^2 dr}{\hat{V}}\; \phi_r^2\left( \frac{r}{L} \right)^{-2\eta}
\nonumber \\
&& 
\sim \frac{\phi_r^2}{1-2\eta/3}
\eea
where we define $\hat{V} = 4\pi L^3/3$ (this differs from eq.(\ref{norm0}).
This leads to the normalization integrals:
\bea
\label{norm10}
&& 
{\cal{N}}^{(n)}=\int \frac{ d^3r}{\hat{V}}\; \phi^n
= \frac{\phi_r^n}{(1-n\eta/3)}
\eea
In eq.(\ref{actionr}) we have the combined action for 
$\chi$ and $\phi$:
\bea
 \label{action5}
 S\!\!&=&\!\!\!\!\int \!\! d^4X\frac{d^3r}{\hat{V}} \!\left( 
 | \phi\partial_\mu{\chi}|^2\!\!- \!\!|\chi \nabla_r{\phi }|^2
\!\!\! -V_{0}(r)|\chi\phi |^{2}\right)
\nonumber \\
& &\!\!\!\!\!\!\!\!\!\!\!\!
-g_0\!\!\int \!\!  d^4\!X \frac{d^3r}{\hat{V}}\! [\bar{\psi}_L(\!X\!
+\!r,t)\psi_{R}(\!X\!-\!r,t)]\chi(X)\phi(r)+h.c. \nonumber \\
\eea
To maintain canonical kinetic terms, the $\chi$ 
field and other parameters must then be renormalized as:
\bea
\chi'&=&\sqrt{{\cal{N}}^{(2)}}\chi
\nonumber \\
g'&=&g{\cal{N}}^{(1)}/\sqrt{{\cal{N}}^{(2)}}=g\frac{(1-2\eta/3)^{1/2}}{(1-\eta/3)}\approx g
\nonumber \\
\lambda'&=&\lambda{\cal{N}}^{(4)}/\sqrt{{\cal{N}}^{(2)}}^4=\lambda\frac{(1-2\eta/3)^2}{(1-4\eta/3)}\approx \lambda
\eea
Hence the renormalized action becomes:
\bea
 \label{action6}
 S&=&\!\!\int \!\! d^4X  \left( 
 |\partial_\mu{\chi'}|^2\!\!- \!\!M^2 |\chi'|^2||\right)
\nonumber \\
& &\!\!\!\!\!\!\!\!\!\!\!\!\!\!\!\!\!\!\!
-g\int \!  d^4X [\bar{\psi}_L(X,t)\psi_{R}(X,t)]\chi'(X)+h.c.
+...
\eea
where the ellipsis is series of higher dimension derivative terms by expanding the $g$
term in $r$.  
Note that
in this scheme $\phi$ is dimensionless, while $\chi$ carries dimensions
of mass, and $\Phi = \phi\chi$ has canonical dimensions of mass.  Recall, we 
treat $\phi$ as dimensionless since
we are only considering it classically at present and it is a static field. 

The energy of the massless solution, $V_{r}=-\eta/r^2$, is finite with an IR cut-off
and is given by the spatial integral over the stress tensor, $T_{00}$.
Integrating by parts and using the equation of motion for the static massless
solution, e.q.(\ref{eigen}),
we obtain surface terms:
\bea
\int_0^L\!\!\!\! 4\pi \frac{r^2 dr}{\hat{V}}(|\nabla_r\phi|^2\!+\!V_r(r)|\phi|^2)
=\left.\!\! \frac{4\pi r^2}{\hat{V}} \phi^*(r)\nabla_r\phi(r)\right|^{ L}_{0}
\eea
Clearly we have at the origin $\nabla_r\phi(r\rightarrow 0)=0$.  Moreover, with the 
dominant $A(r)^{-\eta}$ solution we have  $\hat{V}\sim L^{3-2\eta}$ and,
\bea
m^2\sim \frac{r^2}{\hat{V}}\phi^*(r)\nabla_r\phi(r) \sim -\eta L^{-2}
\;\;\makebox{as}\;\;
r\rightarrow L
\eea
Hence the mass $\sim 1/L$ and is arbitrarily small in the large $L$  limit
and resides on the surface at large $L$.  

We remark that
this represents the flow of ``scale charge''  onto the bounding surface
in the IR (where $\phi\partial_\mu\phi$ is a Weyl current
in the implementation of scale transformations in general relativity
with scalar fields and non-minimal couplings \cite{HR}).
This is somewhat remniscent of topology in which the Chern current,
whose charge represents a topological field configuration in the bulk in odd-$D$
will produce an anomaly on a bounding surface in even-$(D-1)$.  

\subsection{Positive Infrared (Mass)$^2$}

We presently discuss the origin of mass in the context of the extended LL
solution for a BEH-boson.  This will not be a rigorous treatment, 
but rather a sketch of how we think some mechanisms for mass generation may work.
We will return to this in greater detail elsewhere.

To be a physical and normalizeable solution, we require  an IR cut-off, $L$, 
which in turn implies
a mass, $m\sim 1/L$, for the scalar field.
In order to have a mass the $-1/r^2$ potential must deviate from
the scale invariant form in the IR.
We can add a small IR mass term to the theory by explicitly
modifying $V_{loop}$.
If the potential evolves into the form,
\bea
V_{loop,m}=-\frac{\eta}{r^{2}}+m^{2}
\eea
we see that the $\chi,\phi$ equations of motion for an eigenvalue, $M^2$, 
separate:
\bea
\label{eqmo2}
&&\partial^2 \chi(X) =-M^2\chi(X),
\\
&&\left( \frac{d^{2}}{dr^{2}}+\frac{2}{r}\frac{d}{dr}+\frac{\eta}{r^{2}}-m^{2}\right) \phi
(r)=-M^{2}\phi (r),
\eea
and now the solution has mass $M^2=m^2$:
\bea
\Phi=\chi(X)\phi(r)= \phi_r\left( \frac{r}{L}\right)^{-\eta} e^{imt}
\eea
However, this is energetically disfavored for $r>L$
and is no longer a pure eigenstate since we can reduce the
energy by terminating the LL solution and transitioning into pure radiation.
This happens at the scale at which the $m^2$ term dominates the $-\eta/r^2$, approximately
where the potential vanishes:
\beq
m^2 = \eta/L^2, \qquad \; L=\sqrt{\eta}m^{-1}.
\eeq
Hence, at larger distances, $r>>L$ we expect to have a pure radiative solution, as in Region III
of the warm-up barrier potential. 

The mass term in principle can be generated by the running of $g$, or equivalently, $\eta$.
The potential will evolve by the RG as:
\bea
\label{finalpot}
-\frac{\eta}{r^2} \rightarrow -\frac{\eta}{r^2}+\frac{\eta' m}{r}+ m^2
\eea
where $\eta$ develops approximate power law behavior via the RG equation.
Eq.(\ref{finalpot}) is known as a ``Mie potential'' \cite{Mie} and dates from  early days of molecular physics.

This form would imply a large beta function,  $\beta(g)/g \sim n$,  passing through
integer values. This represents a large trace anomaly,
and is the analogue behavior that is seen in the Coleman-Weinberg
potential \cite{CW} (see \cite{HillCW} for a discussion of the trace-anomaly
in this context).  This can in principle occur with perturbative $g$,
but the full formalism using the improved stress tensor is beyond the  scope of this present paper.

This is ``hand-waving'' at this point and requires a more
detailed analysis to understand the solution at the massless LL transition to positive $m^2$ radiation.
The negative $m^2$ case discussed next seems to be more well-defined 
and directly applicable to the BEH-boson.

\subsection{Negative Infrared (Mass)$^2$: RG Trigger for EW Mass Generation}

\begin{figure}
	\centering
	\includegraphics[width=0.5\textwidth]{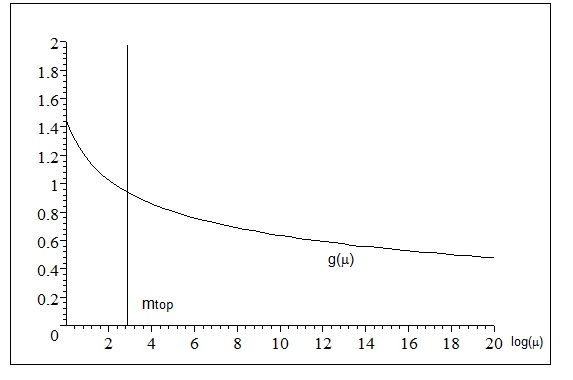}
	\caption{
	RG trajectory of top quark BEH-Yukawa coupling $g$ vs $\log(\mu\;GeV)$ where 
	the vertical line denotes the physical
	top mass, $\log(175\; GeV)$; The gradual rise of the coupling 
	into the infrared (left) is entirely driven by QCD \cite{PR}. 
	}
	\label{fig:loop2}
\end{figure}

The LL solution will terminate at a scale at which 
the $-\eta/r^2$ potential becomes constant $\sim -\eta/L^2$.  
This happens naturally if the Feynman
loop that induces the potential decouples, and
generally requires
that the chiral fermions in the loop acquire a mass, $m_f$.  
The $1/r^2$ potential then ``freezes'' at the scale $L\sim 1/m_f$
and becomes a negative $-m^2\Phi^\dagger\Phi$
term where $m^2 \propto \eta m_f^2$. 
In addition we will have a $\lambda(\Phi^\dagger\Phi)^2/2$
term, induced by fermion loops (see Appendix I).  
Hence, the composite $\Phi$ field will develop a VEV
in the usual way.

In the case of the BEH-boson composed of top and anti-top quarks this would occur
when the top quark acquires a mass.   However, the top quark mass comes from the electroweak symmetry breaking
and the VEV of the BEH-boson.  The formation of the VEV and top quark mass will
then occur in a self consistent way. A consistency condition then determines
the effective value of the running $g$.
While we haven't fully developed the $SU(2)\times U(1)$
isodoublet BEH boson, 
we can get an idea of how this might work for electroweak symmetry breaking 
within our present understanding of the composite system described here.
We presently assume that the BEH boson is composed of top and anti-top quarks
\cite{CTH,Yama,BHL}.  

The vacuum loop potential, in terms of the physical separation
of the constituents, $\rho = 2r$ is given by
\bear
V_{Loop} =-\frac{4\eta}{\rho^2}= -\frac{g^2N_c}{8\pi^2\rho^2}
\eear
$g^2$ evolves by the RG  equation  
as a running in length scale $\rho$. For the top quark this is \cite{PR},
\bea
16\pi^2 \frac{\partial g}{\partial \ln(\rho)}=- g\left(\frac{9}{2}g^2-8g_{QCD}^2\right)
\eea
The solution shows $g$ gradually increasing at large distances, $\rho = \mu^{-1}$,
due to the effects of QCD, which cause it 
to be slightly asymptotically free as seen in Fig.(3). 

However, suppose that the running  $\eta=g^2N_c/32\pi^2$ halts
at some scale $\rho_0$ due to the ``freezing.'' We then have the potential for $\Phi$:
\bea
V=-\frac{4\eta}{\rho_0^2}|\Phi|^2 + \frac{\lambda}{2}|\Phi|^4
\eea
We expect the normalizations of the $g$ and $\lambda$
are not far from the their standard model values, as seen
by the normalization discussion above. In the standard model we have the phenomenological
values, $\lambda \approx 1/4$ and $g=g_{top}\approx 1$.
Hence the fermion loop freezing will lead
to  a spontaneous breaking and $\Phi$ develops a VEV:
\beq
\langle |\Phi|^2 \rangle ={4\eta}/\rho_0^2{\lambda}
\eeq
In principle this can happen for any composite field
given the negative mass term. 
However, for the BEH-boson 
this in turn implies that the top quark develops a mass given by
\bea
m_{top} =  g\langle |\Phi| \rangle= 2g\sqrt{\eta}/\rho_0\sqrt{\lambda}
\eea
Hence spontaneous symmetry breaking 
happens if a consistency condition for $g$ is fullfilled:
\beq
m_{top}^2 \rho_0^2 \approx 4g^2\eta/\lambda \approx \frac{g^4N_c}{2\pi^2}
\eeq
We might expect a cut-off when $\rho_0$ is of order a half wavelength:
\bea
\rho_0 \sim 1/2m_{top}
\eea
Therefore we find:
\bea
g=g_c\approx(\pi^2/6)^{1/4}\approx  1.13
\eea
slightly
larger than the known experimental value, $g=1$.

Our crude result indicates that $g\sim O(1)$ can trigger
a symmetry breaking mechanism that normalizes the LL solution.  There
 will be enhancements by t-channel gluon and $Z$ exchange that tend
to reduce the requisite $g^2$.  Moreover, the fermion loop diagram with insertion of
$m_{top}$ and a single $\phi(r)$ is expected to generate a $\sim gm/r$ Coulomb
interaction in addition to the $\eta/r^2$ potential, so the system
is expected to be described by a Mie potential \cite{Mie}, and 
we again expect a further corrections 
tending to reduce the value of $g$.
We will return this problem in greater detail elsewhere.


\section{Conclusions}

In summary, we began by formulating the bound state problem for a pair of chiral fermions 
in a bosonized
wave-function that represents an s-wave, which can be either a bound state or a pair radiation field.
This is a convenient way to describe the $s$-wave sector,
with the correlated colors, spins and flavor quantum numbers, then
requiring using only
complex scalars.

As an example of the method, we first considered a  short distance dynamics  
that produced a  localized ground state wave-function in a 
simple non-confining barrier potential, $V_r(r)$.
This leads to an approximate, quasi-stable, eigenstate, since it can
decay to free unbound fermions, and we estimated semiclassically the induced Yukawa coupling $g$
to free fermions.

The main point of this paper is that the
presence of the Yukawa coupling to free chiral fermions nontrivially affects the vacuum. 
The coupling, via a Feynman loop, induces
a scale invariant potential, 
$V_{Loop}(r)=-\frac{\eta}{r^2}$.  This
acts upon the valence wave-function. 
This is the quantum loop effect normally considered
in momentum space for the Nambu--Jona-Lasinio model, where it subtracts
from the bound state mass.   In the present case we obtain
the potential coefficient,
$\eta = N_c g^2/32\pi^2$, computed explicitly in  Appendix I.

This  potential, which is
approximately scale invariant, causes the bound state
solution to be of the Landau-Lifshitz
form external to the core \cite{LL}.
We discovered that the LL solutions have a natural affinity with the NJL model.
We see that the NJL critical value of the coupling,
defined by $N_c g^2/8\pi^2=1$
is identical to the critical value defined by the  LL solution, 
i.e., $4\eta  = 1$. 
It is remarkable that the classical LL solutions anticipate the
critical coupling of the NJL model,
where the latter is obtained by a loop calculation. This kind of classical--quantum
correspondence is remniscent of topological solutions, 
Chern-Simons terms and anomalies.

The wave-function solutions in this potential
were first studied by Landau and Lifshitz 
in nonrelativistic quantum mechanics \cite{LL}. 
When these LL solutions are present, boundary
condition matching to the short distance solution will always
force an overall massless scalar bound state solution.  
This requires no fine--tuning and is inherently perturbative.
The pure scale invariant 
potential therefore implies these massless solutions, which we term the ``shroud.''
These become the exterior of the ground state for any
system containing the chiral fermions in a non-confining potential. 
Here the  RG running of $g^{2}$  is soft and can be treated
approximately as a fixed point \cite{PR}.

This behavior appears to be general property of non-confining potential solutions 
with chiral fermions.
Landau and Lifshitz commented on this in the context
of quantum mechanics in their textbook, and apart from negative energy
states (for us, $-M^2$ at short distances, that must be excluded), 
the zero energy ground state is always present.

Far infrared scale breaking can
terminate the LL wave-function, makes the LL solution
normalizeable and is associated with a naturally small
mass for the solution.   We think that the {\em a priori} non-normalizeability of
these solutions may have previously blocked their application.
Presently we have natural IR cut-off, i.e., the BEH mass scale
natually solves the normalization problem, and in this scheme the BEH boson
resembles an inflated balloon.
Scale breaking can come: (a) explicitly; (b) via a Coleman Weinberg mechanism
(which involves the RG evolution of $\lambda$), or
 (c) from a new mechanism involving the 
the RG evolution of the Yukawa coupling $g$.
A  negative $m ^{2}$ appears to arise naturally in the IR. 

This mechanism can therefore yield, with no fine-tuning, a low mass bound state
and an arbitrarily large hierarchy dynamically generated between its core and its mass.  
For the BEH-boson composed of a top quark pair, the RG evolution
is a slowly increasing $g$ in approaching the IR, due to QCD.
We argue that this may be the trigger mechanism for the electroweak scale, the BEH-boson
mass and the top quark mass as one unified phenomenon.  This happens for a
perturbative value of the BEH-Yukawa coupling to top  of ${\cal{O}}(1)$.  
A crude calculation gives $g\approx(\pi^2/6)^{1/4}\approx  1.13$,
compared to $1.0$ experimentally.  While this is in need of further elaboration, which 
we will pursue elsewhere,  optimistically we may be able to precisely predict
the electroweak scale.

Are there any loop-holes in the arguments we have presented?
The massless ground state relies on the deformation of the core wave-function,
and we believe that is a general phenomenon (as did Landau and Lifshitz).  
One might posit a pointlike fundamental scalar boson with fixed, nondeformable mass, as
in the usual momentum space formulation of the NJL model.
In this case the shroud solution would apparently
not exist and the exterior is radiation.   We believe this is an ambiguity
of the NJL model in that it doesn't specify a short distance dynamics and
is treated only as a cut-off theory in momentum space.  A better specification
of the NJL model is therefore required in configuration space, and it may be
that the shroud solution is mandated, since we have seen the equivalence of the critical 
behavior of the LL solution and the NJL model.  Can we construct a
non-deformable core solution which would forbid the matching to the LL solution?
How does the dynamical symmetry breaking in the NJL model manifest itself through
the LL supercritical, $\eta >1/4$, solution?
A pair of
chiral fermions bound into a black hole \cite{HBH} poses an intriguing problem
problem along these lines.  These are questions to be addressed.

We are also
relying on the non-existence of negative $M^2$ states at short distance {\em when
the vacuum loop potential is included.} If such solutions exist then the chiral
symmetry is spontaneously broken at short distance 
and the chiral fermions acquire mass of order $M$.
This would be a disaster for any composite BEH-boson scenario.  We have not 
formally proven that
we can always exclude such solutions, but we know that negative energy
bound states in spherical potentials
are restricted and often do not exist in weak coupling.  Ref.\cite{LL} assert that no 
such negative energy states exist when $-\beta/r^2$ fills all of space and one has weak coupling.
This is realized in the case of barrier potentials, together
with $\eta < 1/4$.
Hence, we believe there is likely
a large, non-fine-tuned range of parameters over which the massless ground state exists with
no negative $M^2$ solutions at short distances.

We have only treated $\phi$ classically at present and we are able
to normalize it as a dimensionless field.  It is a static configuration and it
is not subject to canonical normalization, though it enters the path integral
and would presumeably be integrated (perhaps in analogy to instantons). 
Classically it satisfies a static differential equation that generates the LL solution.
We haven't investigated fully the conceptual issues associated with the 
composite field factorization or path integration over $\phi$.

There is much to do to further develop and test and apply this theory. 
For example, the extension to the many 
flavors of the standard model requires some kind of novel
interactions, suggestive of something akin to extended technicolor interactions \cite{ETC}.
The softness
of the BEH-boson above the threshold implies a 
significant and potentially observable, non-pointlike form factor.
This may be probed in sensitive measurements of decay modes and coupling constants.
It may be optimally
probed in a machine such as a muon collider, a  BEH-factory with s-channel production of
the BEH-boson \cite{muon}.

It is possible that there are many low mass scalar bound states of the chiral standard model
fermions, perhaps due to gravitation.  Hence a scalar democracy consistng
of low mass $s$-wave combinations of all SM fermion pairs
may exist \cite{HMTT}. This possibility is
experimentally accessible at LHC upgrades, searching for the $b\bar{b}$
combination \cite{HP}.

We recently pointed out that mini-blackholes are expected to form near $M_{Planck}$
composed of any pair of chiral fermions with the quantum numbers of
the BEH-boson.  We argued that they may be very light due
to unknown dynamics, appealing to the existence BEH-boson as evidence \cite{HBH}.
Here we offer the present mechanism to further substantiate this claim. It
may be interesting to study the LL solutions and shroud 
surrounding a mini--Reissner-Nordstrom black hole.

We remark that the  top quark does not necessarily have to be the constituent of the BEH boson
even in the context of the present scheme.
It is alternatively  possible that we have ``BEH as a PNGB'' theory \cite{Georgi}
(e.g., a ``Little Higgs theory''), where the parent scalar is an shroud solution
to solve the large hierarchy problem, and spontaneously broken chiral symmetry
solves the little hierarchy problem.  Hence we think the shroud solution offers an arsenal
of model building tools, perhaps even a naturalization of traditional grand unified theories
sucha s $SU(5)$ or $SO(10)$.

While this is a candidate mechanism that may provide a 
solution to the gauge hiearchy problem and a
natural low mass  BEH-boson,
it may also be partially operant within QCD and account for the unexpectedly low
mass of the $\sigma$-meson.
The $\sigma $-meson of QCD appears at a surprisingly lower mass scale, 
$\sim 500$  MeV, rather than the expected $\sim 1$ GeV (this is the $f_0(500)$
and for discussion see \cite{Crewther}).
Since it is $m^2$ that matters here, this seems to be a mini-hierarchy
of order $\sim 1/4$.  This may be a ``mini-shroud effect" extending 
from the expected
scale $\sim 1$ GeV, to the observed mass, $\sim 500$ MeV, 
and would be expected in the context of a chiral constituent quark model.

Therefore, it is our conclusion that composite low mass scalars composed of chiral fermions can exist
naturally.  The ``custodial symmetry'' is scale invariance together with chirality,
acting within the internal wave-functions
and dynamically realizing the  approximate masslessness. This
suggests the BEH-boson is composed perturbatively of top and anti-top quarks,
and the BEH boson is an extended object, of order $\sim 1/m_{top}$ in scale,
behaving coherently as a pointlike state in current processes at current LHC energies.
This may suggest a rich spectroscopy of other flavor combinations in bound states.
We  hope to 
devise ways of testing this in 
foreseeable experiments.  
\newpage

\appendix

\section{Calculation of the Vacuum Loop Potential}

\begin{figure}
	\centering
	\includegraphics[width=0.20\textwidth]{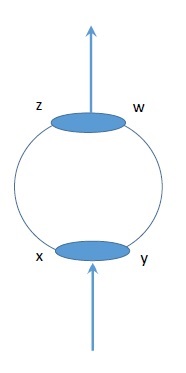}
	\caption{ Loop with wave-function vertices.} 
	\label{fig3:loop}
\end{figure}

 \subsection{Pointlike Limit}

We consider the bilocal action of eq.(\ref{actionr}):
\bea
 \label{actiont}
S\!=&&\!\!\!\!\!\!\!\!\int \!\!\frac{ d^3 r}{\hat{V}} d^4\!X\! \left( 
  | \phi^2||\partial_\chi{\chi}|^2\!\!- \!\!|\chi|^2|| \partial_r{\phi }|^2
\!\!\! -\!V_{r}(\vec{r})|\chi^2||\phi |^{2}\right)
\nonumber \\
& &\!\!\!\!\!\!\!\!\!
-\;g\!\!\int \!\!  \frac{ d^3r}{\hat{V}} d^4X [\bar{\psi}_L(\vec{X}+{\vec{r}})
\psi_{R}(\vec{X}-{\vec{r}})]\chi(\vec{X})\phi(\vec{r})
\nonumber \\
&& \!\!\!\!\! \makebox{where,} \;\;\;\;\;
\vec{x} =\vec{X}+\vec{r}, \;\;\;\;\; \vec{y} =\vec{X}-\vec{r}.
\!\!\!\!\!\!\!\!\!\!\!\!
\eea
where we have set $z=1$ for simplicity.

To test the composite action,
we compute the effective potential that is induced for
the field
$\chi(X)$ by the fermions, for a point-like bound state, $\phi\sim \delta^3(r)$. 
Assume that we have a short-distance solution of the $\phi$ static
spatial equation:
\bea
\vec\nabla_{r}^2\phi(\vec{r})-V(\vec{r})\phi(\vec{r})=M^2\phi(\vec{r})
\eea
where $M^2$ is the eigenvalue, as in our discussion of the barrier potential.\footnote{Alternatively
we could take a simple harmonic oscillator potential bounded by $R$ as $V=\kappa {r}^2\theta(R-r)$
which has a Region I Gaussian solution, and a Region II steady state radiation field.  This allows a
straightforward pointlike limit where the Gaussian becomes $\sim \delta^3(\vec{r})$.}
We then take a limit in which $\phi\sim \delta^3(\vec{r})$, and
define the pointlike dimensionless field:
\bea
\label{A6}
&& \!\!\!\!\!\!\!\! \phi(\vec{r})\rightarrow {\cal{N}}\phi_0{\hat{V}}
\delta^3(\vec{r}),
\nonumber \\
&&\!\!\!\!\!\!\!\!\makebox{hence,}
\int \frac{d^3r}{\hat{V}} |\phi|^2 ={\cal{N}}^2|\phi_0|^2\hat{V}\delta^3_r(0)=1
\nonumber \\
&&\!\!\!\!\!\!\!\!\makebox{and,}
\int \frac{d^3r}{\hat{V}}|\phi| ={\cal{N}}|\phi_0|  =1
\eea
where ${\cal{N}}^{-1}=|\phi_0|$, and we define $\hat{V}^{-1}=\delta^3_{\vec{r}}(0)$.
Then the action becomes,
\bea
\label{action4}
 S'&=&\!\!\!\int \!\! d^4X \big( 
 | \partial_{\mu}\!\chi|^2- M^2|{\chi }|^2\big)
\nonumber \\
& &\!\!\!\!\!\!\!\!\!\!\!\!
-g\!\int \!\!d^4X [\bar{\psi}_L(X)\psi_{R}(X)]\chi(X)+h.c.
\eea
The loop integral could now be done using the action if eq.(\ref{action4})
since, in Fig.(4), $x=y$ and $w=z$ having integrated out the pointlike internal $\phi$ field.
However, it is useful to do the loop integral
from the point of view of the composite field $\phi$ as a warm-up to the non-pointlike case.

First we note that the four vertex variables of Fig.(4) can be written as:
\bea
\vec{r} &=& \half(\vec{x}-\vec{y}),\qquad
\vec{r}\;{}'=\half( \vec{w}-\vec{z}),
\nonumber \\
\vec{X}&=&\half(\vec{x}+\vec{y}),\;\;\;\;
\vec{X}'=\half(\vec{w}+\vec{z}),
\eea
Hence,
\bea
\label{coords}
&&\!\!\!\!\!\!\!\!
\vec{x}-\vec{z}=\vec{r}+\vec{r}\;{}'+\vec{X}-\vec{X'},
\nonumber \\
&&\!\!\!\!\!\!\!\!
\vec{w}-\vec{y}=\vec{r}+\vec{r}\;{}'-\vec{X}+\vec{X'},\;\;\;\;
\eea

Consider the T-ordered product from eq.(\ref{actiont}) (including an $(i)^2$ factor
from $e^{iS}$ and $-1$ from anti-commutation), and notation
$\int_{x...z}=\int d^4x...d^4z$:
\bea
\label{start}
&& 
\!\!\!\!\!\!\!\!(i)^2g_0^2\!
\int_{xywz}\!\!\!\!\!\!\! \!\!\!\langle 0|T\![\bar{\psi}_L(x)\psi_{R}(y)]\!\!
\; [\bar{\psi}_R(w)\psi_{L}(z)]|0\rangle \Phi(x,y)\Phi^\dagger(z,w)
\nonumber \\
&&\!\!\!\!\!\!\!\!
=
g_0^2N_c
\int_{xywz} \!\!\!\!\!\!\!\!\Tr (S_F(x-z)S_F(w-y){\cal{P}}_5)\Phi(x,y)\Phi^\dagger(z,w)
\nonumber \\
&&\!\!\!\!\!\!\!\!
=
g_0^2N_c
\int\!\! \frac{d^3r}{\hat{V}}\frac{d^3r'}{\hat{V}}d^4Xd^4X'\;
\chi(X)\phi(r)\chi(X')^*\phi(r')^*
\nonumber \\
&&\times
\Tr (S_F(\vec{r}\!+\!\vec{r}\;{}'+\!\vec{X}\!-\!\vec{X'}\!)
S_F(\vec{r}\!+\!\vec{r}\;{}'-\!\vec{X}\!+\!\vec{X'}\!){\cal{P}}_5)
\eea
where ${\cal{P}}_5=(1-\gamma^5)/2$, where we included
 $\delta(x^0-y^0)$ and $\delta(z^0-y^0)$ factors for
the single time  gauge fixing, and the volume normalization,
$\int d^4x d^4y \delta(x^0-y^0) \rightarrow \int d^4Xd^3r/\hat{V}$.

Now define $\chi = \chi_0\exp(-iP_\mu X^\mu)$, with 
the pointlike $\phi={\hat{V}}\delta^3(\vec{r})$ as in
eq(\ref{A6}).
We then obtain for eq.(\ref{start}) with
arbitrary in (out) momenta $P$ ($P'$):  
\bea
&&\!\!\!\!\!
=
g_0^2N_c|\chi_0|^2\!\!
\int\!\! d^4Xd^4X' 
\Tr(S_F({X}\!-\!{X'})S_F({X'}\!-\!{X}){\cal{P}}_5)
\nonumber \\
&& \qquad \times e^{-iP_\mu X^\mu}e^{iP'_\mu X'^\mu}
\eea
Note  cancellation of $\hat{V}$ factors.
We now use the momentum space Feynman propagator, 
\bea
S_F(x-z)=\int \frac{d^4\ell}{(2\pi)^4} \frac{i\slash{\ell}}{\ell^2+i\epsilon} e^{i\ell\cdot(x-y)}
\eea
Taking the trace, and omitting a factor of $g_0^2N_c|\chi_0|^2$
which we restore at the end, and integrating over $X$, $X'$, we have:
\bea
&=& \!\!\! - \!\! \int_{XX'}\!\!\frac{d^4\ell}{(2\pi)^4}\!\frac{d^4\ell'}{(2\pi)^4}\!   
\Tr \big({\cal{P}}_5\frac{\slash{\ell}}{l^2}\frac{\slash{\ell}'}{\ell'^2}\big) 
e^{-i\ell\cdot(X-X')}e^{-i\ell'\cdot(X'-X)}
\nonumber \\
&& \times \;
e^{-iP_\mu X^\mu}\;e^{iP'_\mu X'^\mu}
\nonumber \\
&=&-2(2\pi)^4\delta^4(P-P')\int\frac{d^4\ell}{(2\pi)^4}   
\frac{\ell\cdot(\ell+{P})}{\ell^2(\ell+P)^2}
\nonumber \\
&=&-2\int\!\! d^4X\int_0^1dx\int\frac{d^4\hat\ell}{(2\pi)^4}   
\frac{(\hat\ell^2-x(1-x){P}^2)}{(\hat\ell^2+x(1-x)P^2)^2}
\nonumber 
\eea
Here $\hat\ell = \ell-x P$ and we drop terms odd in $\hat \ell$. 
We have identified the $(2\pi)^4\delta^4(P-P')=\int\!\! d^4X$ the volume of space=time.
in the $P=P'$ limit.
We perform a
Wick rotation:  $\hat\ell_0 \rightarrow i\hat\ell_0$ so $d^4\hat\ell\rightarrow i d^4\hat\ell$
and $\hat\ell^2\rightarrow -\hat\ell^2_0-\vec{\hat\ell}^2\equiv -\ell^2$
and $d^4\ell\rightarrow \pi^2 \ell^2d\ell^2$, and we impose a cut-off, $\Lambda$,
hence:
\bea
&&\!\!\!\!\!\!\approx \frac{ig^2N_c}{8\pi^2}|\chi_0|^2\int\!\! d^4X 
\left(\left(\Lambda^2-\mu^2\right)+
\frac{1}{2}P^2\ln(\Lambda^2/\mu^2)\right)
\nonumber \\
\eea
where we restored the $g^2N_c$ factor.
This then enters the action as a potential and a kinetic term
in the NJL model following \cite{BHL,CTH},
upon restoring $g^2N_c$, 
\bea
V&=& -\frac{g^2N_c}{8\pi^2}\left(\Lambda^2-\mu^2\right)|\chi|^2
\nonumber \\
K &=&
\frac{g^2N_c}{16\pi^2}\!\ln(\Lambda^2/\mu^2) \partial_\mu\chi^\dagger \partial_\mu\chi
\eea

\subsection{Extended Composite Limit}

We are now interested in the non-pointlike composite model.
We first require  the potential energy as a function
of an arbitrary internal field configuration $\phi(r)$ for a particular
value of $r$.  

This is analogous to the Coleman-Weinberg potential, where
we would be interested in the potential energy when 
the VEV of a field $\phi$ is constrained to a particular value
$\phi_0$. 
In Schroedinger picture this corresponds to a vacuum wave-functional,
$\Psi(\phi)$, where $\int D\phi \Psi^*(\phi)\phi\Psi(\phi)=\phi_0$.
To obtain the potenial we compute the expectation of
the Hamiltonian by integrating over the fluctuations in $\phi$ subject to this constraint
and minimizing wrt all other parameters in $\Psi$.
From  a path integral point of view we start on a time
slice $t=-\infty$ in which $\langle\phi\rangle=\phi_0$, integrate over all space-time
fluctuations in $\phi$ and
end on $t=\infty$ with $\langle\phi\rangle=\phi_0$. Typically the field
VEV is obtained by addition of a source, $J\phi$ 
followed by a Legendre transformation to the shifted field
(The source cancels linear terms in $\phi_0$).
Then $i\times$(the log of the path integral)
is  the effective potential as a function of $\phi_0$.

Note that in our present problem we have
four space-time vertices, $(x,y,z,w)$ s in Fig.(4).
We fix the single time gauge with insertion into the integrand 
of $\delta(x^0-y^0)\delta(w^0-z^0)$.
We implement the fixed $r$ constraint by inserting
a $\hat{V}\delta^3(\vec{r}-\vec{r}\;{}')$ into our integrand, with
 the two vertices, $\phi(\vec{r})$, $\phi(\vec{r}\;{}')$.
 
The loop integral of Fig.(4), as in eqs.(\ref{coords},\ref{start}), becomes, 
\bea
&&\!\!\!\!\!
=
g_0^2N_c|\chi_0|^2
\int \frac{d^3r}{\hat{V}}\frac{d^3r'}{\hat{V}}d^4Xd^4X' 
\nonumber \\
&&
\times
\Tr \big(S_F(\vec{r}\!+\!\vec{r}\;{}'+\!\vec{X}\!-\!\vec{X'}\!)
S_F(\vec{r}\!+\!\vec{r}\;{}'-\!\vec{X}\!+\!\vec{X'}\!){\cal{P}}_5\big)
\nonumber \\
&&\;\;\;\;\;\times \phi(\vec{r})\phi(\vec{r}\;{}')^\dagger e^{-iP_\mu X^\mu}e^{iP'_\mu X'^\mu}
\;\hat{V}\delta^3(r-r')
\nonumber \\
&&\!\!\!\!\!
=-F\!\!\int\frac{d^3r}{\hat{V}}\frac{d^4\ell}{(2\pi)^4}\frac{d^4\ell'}{(2\pi)^4}   
\Tr{\cal{P}}_5 \frac{\slash{\ell}}{l^2}\frac{\slash{\ell}'}{\ell'^2} 
\nonumber \\
&& 
\qquad \times
|\phi(\vec{r})|^2e^{2i\vec{\ell}\cdot\vec{r}}
e^{2i\vec{\ell'}\cdot\vec{r}}(2\pi)^4\delta^4({\ell}-{\ell}'-{P})
\eea
Here  we performed the $x^0,y^0,w^0$ and $z^0$ time integrals,
and,
\bea
&&\!\!\!\!\! \!\!\!\!\! \!\!\!\!\!\!\!\!\!\! F\!=\!
g^2\!N_c|\chi_0|^2(2\pi)^4\delta^4(P\!-\!P')\!
=\! g^2N_c\!\!\int\!\! d^4X|\chi_0|^2.
\eea
We treat $P, P'$ as pure timelike (ie, $\vec{P}\cdot \vec{x}=0$, etc.),
do the $\ell'$ integral, and take the trace:
\bea
&&
=2F\int \frac{d^3r}{\hat{V}}\frac{d^4\ell}{(2\pi)^4}   
\frac{\ell\cdot(\ell+P)}{\ell^2(\ell+P)^2} 
|\phi(\vec{r})|^2e^{4i\vec{\ell}\cdot\vec{r}}
\nonumber \\
&&=2F\int_0^1dx\int \frac{d^3r}{\hat{V}}\frac{d^4\hat\ell}{(2\pi)^4}   
\frac{(\hat\ell^2-x(1-x){P}^2)}{(\hat\ell^2+x(1-x)P^2)^2}
|\phi(\vec{r})|^2e^{4i\hat{\vec{\ell}}\cdot\vec{r}}
\nonumber \\
&&\approx F\!\!\int\!\! \frac{d^4\hat\ell}{(2\pi)^4} \frac{d^3r}{\hat{V}}  
\left[\frac{2}{\hat\ell^2}-
\frac{P^2}{\hat\ell^4}+...\right]
|\phi(\vec{r})|^2e^{4i\hat{\vec{\ell}}\cdot\vec{r}}
\eea
where
$\hat\ell = \ell-x P$.
Now we don't Wick rotate, and do the $\ell_0$ integral by
residues.  We have:
\bea
&&\int\frac{d^4\ell}{(2\pi)^4}\frac{1}{\ell^2-\mu^2+i\epsilon}
=
\frac{i}{2}\int\frac{d^3\vec{\ell}}{(2\pi)^3}\frac{1}{(\vec{\ell}^2+\mu^2)^{1/2}}
\nonumber\\
\eea

We perform the $\ell_0$ integrals and then the 
polar angle integrals:
\bea
&&=\frac{i}{2}\int\frac{d^3r}{\hat{V}}\frac{d^3\vec{\ell}}{(2\pi)^3}   
\left[\frac{2}{|\vec\ell|}+
\frac{P^2}{2|\vec\ell|^3}\right]
|\phi(\vec{r})|^2e^{4i\vec{\ell}\cdot\vec{r}}
\nonumber\\
&&=i\int \frac{d^3r}{\hat{V}}\int^{\Lambda}_\mu\frac{2\pi d|\vec\ell|}{(2\pi)^3}   
\left[2+
\frac{P^2}{2|\vec\ell|^2}\right]
|\phi(\vec{r})|^2\frac{\sin(4|\vec{\ell}||\vec{r}|)}{4|\vec{r}| }
\nonumber\\
\eea
and upon restoring overall factors we have
the result:
\bea
\label{resultf}
&&=ig^2N_c\!\!\int \! \! d^4X|\chi_0|^2\int \frac{d^3r}{\hat{V}} \frac{|\phi(\vec{r})|^2}{8\pi^2}\times 
\nonumber\\
&&\!\!\!\!\!\!\!
\times\left[\frac{1}{4|\vec{r}|^2}(\cos(4\mu |\vec{r}|)-[\cos(4\Lambda |\vec{r}|)])\!\right.
\nonumber \\
&&\left.
\!\!\!\!\!+\!\frac{P^2}{2}\!\left(\frac{\sin(4\mu |\vec{r}|)}{2\mu |\vec{r}|}
- 2\gamma-\ln(16\mu^2 |\vec{r}|^2)\right)
\right]
\eea
using 
\bea
&&\!\!\!\!\!\!\!\!
\int_\mu^\Lambda \sin(2xR)dx =\frac{\cos2\mu R-[\cos2\Lambda R ]}{2R}
\nonumber\\
&&\!\!\!\!\!\!\!\!
\frac{1}{2}\int_\mu^\Lambda \frac{\sin(2xR)}{x^2}dx 
\nonumber \\
&&
\approx 
\frac{\sin(2\mu R)}{2\mu}\!-\!R(\gamma +\ln(2\mu R))+\!O\left(\!\frac{1}{\Lambda}\right)
\nonumber \\
\eea
and we drop the rapidly oscillating terms such as $\cos(2\Lambda r) $.

Now we assume small $\mu r$, i.e,. separation between the valence fermions
smaller than the IR cut-off $\mu^{-1}$.
Restoring an overall factor of $g^2N_c$ we see that eq.(\ref{resultf})
leads to the vacuum loop potential:
\bea
\label{resultm}
&&\!\!\!\!\!\!=ig^2N_c\!\!\int \!\!\!d^4X
|\chi_0|^2\int\!\!\frac{d^3r}{\hat{V}} \frac{|\phi(\vec{r})|^2}{8\pi^2}
\left[\frac{(\cos(4\mu |\vec{r}|)-[\cos(4\Lambda |\vec{r}|)]}{4|\vec{r}|^2}\!\right]
\nonumber \\
&&\;\;\;\;\rightarrow i \int d^4X |\chi(X)|^2\int\frac{d^3r}{\hat{V}}\; 
\frac{g^2N_c\;|\phi(\vec{r})|^2}{32\pi^2\;|\vec{r}|^2}
\eea
where $\cos(\Lambda r)$ oscillates rapidly and averages to zero 
for small fluctuations in $r$, and we drop it.

Eq.(\ref{resultm})
is our main result, corresponding to $i\times$(action)
and we see the sign in the action is positive, denoting an
attractive potential:
\bear
V_{loop}= -\eta/r^2\qquad   \eta = \frac{g^2N_c}{32\pi^2}
\eear
where we have renormalized the kinetic terms $Z_\chi \rightarrow 1$.

Note the behavior of the kinetic term in eq.(\ref{resultf}) :
\bea
&&\!\!\!\!\! \!\!\!\!\! \!\!\!\!\! 
\rightarrow\;
i\int d^4X |\partial \chi|^2 \int\frac{d^3r}{\hat{V}} \frac{g^2N_c|\phi(\vec{r})|^2}{16\pi^2}
\nonumber \\
&& \qquad \times \left(2
- 2\gamma-\ln(16\mu^2r^2)\right)
\eea
We see that the coefficient and argument of the log matches the
logarithmic running in the Nambu-Jona-Lasinio model as 
in eq.(\ref{five}), with $4\mu^2r^2 \sim \mu^2/M^2$
\bea
&&\!\!\!\!\! \!\!\!\!\! \!\!\!\!\! 
\rightarrow\;
i \frac{g^2N_c}{16\pi^2}\int d^4X |\partial \chi|^2\big(c
+\ln(\Lambda^2/r)^2\big) 
\eea
using the normalization, eq.(\ref{norm10}) and to order $g^2$.
This indicates that the logarithmic RG running of renormalized couplings
in the variable $\ln(r)$ will be  given consistently with full RG equations.

 \subsection{Quartic Interaction}
 
 As in the NJL model, the fermion loops will induce a quartic interaction. By the
 scale symmetry of the factorized bilocal field, we will have a term in the action
 \bea
 -\frac{\lambda}{2}\int d^4X \frac{d^3\vec{r} }{\hat{V}} (\chi^*\chi)^2(\phi^*\phi)^2=
 \eea
 We can infer from the previous calculations that the loop will have
 four bilocal vertices and takes the form:
\bea
&&\lambda = 2g^4N_c\!\!\int d^4X|\chi_0|^2\int \frac{d^4\hat\ell}{(2\pi)^4}\frac{d^3r}{\hat{V}}   \;
\frac{1}{\hat\ell^4}
|\phi(\vec{r})|^4e^{8i\hat{\vec{\ell}}\cdot\vec{r}}
\nonumber \\
&&=2ig^4N_c\!\!\int d^4X|\chi_0|^2\int^{\Lambda}_\mu \frac{d^3r}{\hat{V}}\frac{2\pi d|\vec\ell|}{(2\pi)^3}   
\frac{|\phi(\vec{r})|^4}{2|\vec\ell|^2}
\frac{\sin(8\vec{\ell}||\vec{r}|)}{8|\vec{r}| }
\nonumber\\
&&=2ig^4N_c\!\!\int d^4X|\chi_0|^2\int\frac{d^3r}{\hat{V}} \frac{|\phi(\vec{r})|^4}{8\pi^2}\times 
\nonumber\\
&&\qquad \qquad\times
\left(\frac{\sin(8\mu |\vec{r}|)}{8\mu |\vec{r}|}
- \gamma-\ln(8\mu |\vec{r}|)\right)
\eea
The log evolution matches the result for the pointlike case with $4\mu^2r^2 \sim \mu^2/M^2$.
\bea
\lambda =c_{2}+\frac{2N_{c}g^{4}}{16\pi ^{2}}%
\ln \left(\! \frac{M^2}{\mu^2 }\!\right).
\eea

\vspace{0.5in}

\section*{Acknowledgments}
I thank Bogdan Dobrescu for discussions, and
the  Fermi Research Alliance, LLC under Contract No.~DE-AC02-07CH11359 
with the U.S.~Department of Energy, 
Office of Science, Office of High Energy Physics.

\end{document}